# Adhesive forces inhibit underwater contact formation for a soft-hard collision

Mengyue Sun[1], Nityanshu Kumar[1], Ali Dhinojwala[*], Hunter King[*]

*The University of Akron, Department of Polymer Science, Akron, OH - 44325*

[1]M.S. and N.K. contributed equally to the work.
[*] Correspondence should be addressed to ali4@uakron.edu and hking@uakron.edu

Thermodynamics tells us to expect underwater contact between two hydrophobic surfaces to result in stronger adhesion compared to two hydrophilic surfaces. However, presence of water changes not only energetics, but also the dynamic process of reaching a final state, which couples solid deformation and liquid evacuation. These dynamics can create challenges for achieving strong underwater adhesion/friction, which affects diverse fields including soft robotics, bio-locomotion and tire traction. Closer investigation, requiring sufficiently precise resolution of film evacuation while simultaneously controlling surface wettability has been lacking. We perform high resolution in-situ frustrated total internal reflection imaging to track underwater contact evolution between soft-elastic hemispheres of varying stiffness and smooth-hard surfaces of varying wettability. Surprisingly, we find exponential rate of water evacuation from hydrophobic-hydrophobic (adhesive) contact is 3 orders of magnitude lower than that from hydrophobic-hydrophilic (non-adhesive) contact. The trend of decreasing rate with decreasing wettability of glass sharply changes about a point where thermodynamic adhesion crosses zero, suggesting a transition in mode of evacuation, which is illuminated by 3-Dimensional spatiotemporal heightmaps. Adhesive contact is characterized by the early localization of sealed puddles, whereas non-adhesive contact remains smooth, with film-wise evacuation from one central puddle. Measurements with a human thumb and alternatively hydrophobic/hydrophilic glass surface demonstrate practical consequences of the same dynamics: adhesive interactions cause instability in valleys and lead to a state of more trapped water and less intimate solid-solid contact. These findings offer new interpretation of patterned texture seen in underwater biolocomotive adaptations, as well as insight toward technological implementation.

adhesion | soft-hard contact | biological contact | underwater contact mechanics

**Significance**
Understanding underwater contact mechanics between soft materials and hard surfaces is of prime importance due to its ubiquity and practical relevance. From biological creatures climbing on flooded surfaces and human figures getting grip on wet surfaces to adhesives sealing a wound underwater, performance depends upon water evacuation to achieve contact. We report a counter-intuitive relationship between adhesion and water evacuation. Surfaces with stronger thermodynamic work of adhesion show slower evacuation rates as a result of water entrapment in isolated puddles. Interestingly, the evacuation rates speed up at the point of zero underwater work of adhesion. The insights pave the way for developing materials for achieving underwater adhesion/grip and avoiding liquid entrapments via proper combination of chemistry and surface structure.

**F**rom wet tire traction to underwater adhesion, the evacuation of a thin layer of water between two surfaces is essential for contact formation and subsequent interactions (1-3). Elegant examples in nature use many different strategies to adhere and grip strongly. For example, biological adhesive systems such as mussel foot protein has demonstrated the need to displace interfacial water and the role of hydrophobicity in achieving high underwater adhesion (4-6). On the other hand, spider silk uses hydroscopic compounds in aggregate glues to get rid of interfacial water and hence hunt in wet habitats (7). Apart from solely chemistry-based solutions, the nature has also employed the surface patterning to remove interfacial water and, hence, achieve high underwater adhesion and friction, for example in the case of tree-frogs and geckos (8-11). They rely on weaker van der Waals interactions and water evacuation to create a dry adhesive contact. Additionally for tree-frogs, one finds features such as patterned toepads that have channels designed to collect the water during the attachment cycle and keep the contact interface dry (8, 12), reminiscent of tire tread patterns that help drain the water during rolling on wet surfaces (12). Despite its ubiquity in biological and everyday phenomena, a practical understanding of the dynamics of evacuating fluid between a soft elastomer and a solid surface during and after collision remains elusive due to the complex interplay between adhesion, lubrication forces, solvation forces, roughness and material mechanics (13-15).

The initial footsteps toward the problem befell eight decades ago by Eirich and Tabor (16). Their simple hydrodynamic calculations show that the pressure mediated by a viscous liquid, between metal surfaces brought into contact can be large enough to cause first elastic, and then plastic deformation. Subsequent work demonstrated that elastohydrodynamic forces also cause an increase in the liquid's viscosity (17). For contacting pairs where one of the surfaces is soft, the large film pressures are not observed, and the liquids entrapped can be considered isoviscous. This case was first studied by Roberts and Tabor in which an optically smooth rubber surface approaching glass substrate was found to deform elastically before contact and entrap the liquid, resulting in a bell-shaped elastic deformation (13, 14). The evacuation process of the entrapped liquid was studied using optical interferometry and a liquid of sufficiently high viscosity was chosen to slow down the evacuation rates so that it could be captured as a function of time. The equilibrium film thicknesses (~ 20 nm) for water were influenced by double-layer forces and a function of electrolyte solution. The evacuation dynamics were then analyzed to estimate the viscosities of the solutions using the squeeze-film equation by Reynolds (18). For the case of distilled water entrapment, Roberts and Tabor highlighted that the Reynolds equation was not applicable due to collapse of thin film and entrapping of water pockets. Later, Davis et al. elegantly integrated the lubrication forces and linear elasticity to formulate the collision theory for soft elastic spheres in a fluid (19).

Several studies have extended the elastohydrodynamic theory to systems in which the soft surface is an elastic film supported by a hard substrate (20-22). There, a significant amount of stress is transferred to the underlying substrate resulting in a higher effective elastic modulus and this reduces the elastohydrodynamic response in comparison to that observed for the elastic-half space made of the same material. Frechette et al. measured the deformation and hydrodynamic forces simultaneously to explain the role of elastic forces on hydrodynamic interactions and hence provided insights into the dynamics of contact formation due to the release of stored elastic energy (23). They also disentangled the contribution of elastic compliance and surface roughness to elastohydrodynamic deformation by comparing the measured film thicknesses with those predicted from lubrication theory (15, 20, 24). The advancement in elastohydrodynamic theory has provided a central understanding about the initial state (before any contact) of systems critical to soft coatings for tribology, adhesives and biomaterials. However, desired end properties such as underwater adhesion or friction between soft-hard contacting pairs require the evacuation of trapped water for intimate solid-solid contact, and this condition is still difficult to comment upon without a detailed understanding of the processes following the initial deformation, especially when long-range, adhesive interactions are not negligible.

Here we report the evolution of underwater contact between soft elastomeric lens and hard surface of variable surface energy by visualizing the initial contact formation and measuring the evolution of the entrapped liquid film over time. We determine the dependence of fluid evacuation on the surface wettability and elastomer modulus in the underwater collision as it is essential to any observed transient and final adhesion and friction state. Our results have implications for problems in which the timescale of wet contact is relevant: from tire traction on wet roads to design of underwater adhesives and lubricants to understanding biological solutions for wet traction and adhesion.

**Experimental Setup.** 3D imaging of the contact made between a soft elastomer and hard substrate underwater was achieved using the principle of frustrated total internal reflection (FTIR) (details in Methods). Fig. 1A shows a sketch of the experimental setup where a BK7 glass (refractive index $\mu = 1.52$) prism has the top surface enclosed in a polytetrafluoroethylene chamber filled with water ($\mu = 1.33$). The polydimethylsiloxane (PDMS, $\mu = 1.43$) was selected as an elastomer because of several reasons. The PDMS has a negligible swelling ratio in water, which ensures constant material properties and reduces the possibility of absorption of any water throughout the observed process (25). The PDMS elastomeric network is fully crosslinked and does not contain any nanoparticles for reinforcement. Additionally, the PDMS lenses (blue hemisphere in Fig. 1A) used were all Soxhlet extracted (details in SI, Section 1) for three days before use to ensure the removal of uncross-linked chains which can potentially change the viscosity and wetting of entrapped water, hence the evacuation dynamics. The PDMS lens is brought into contact with glass substrate with a translational Z-stage connected to a stepper motor, at a constant approach velocity sufficient to cause elastohydrodynamic deformation of the elastomer (0.9 mm/sec). Images are captured by high-speed camera at different frame rates based on the time span of the process studied.

We explored the parameter space encompassing three moduli (0.7, 1.5 and 9.7 MPa) for soft PDMS lens and six substrates with water contact angle in air ranging from 0° to 108°. The PDMS lenses of variable elastic modulus were synthesized by employing network theory

(details in SI, Section 1) (26). Evolution of contact area of the lenses while pressed against a low surface energy (~ 24 mJ/m$^2$) monolayer (octadecyltrichlorosilane (OTS) treated silicon wafer) was analyzed to extract the modulus using the JKR model (details in SI, Section 2). To prepare substrates with varying contact angles, an OTS monolayer was solution deposited on glass prism following the procedure described in the literature (3). The static water contact angle on OTS-*t*-glass prism is measured as 108° ± 3° (3 repeats at different spots). Lower contact angles (87.7°± 0.3°, 51.2°± 1.8°, 35.9°± 1.0°, 25.4° ± 0.8°, 0° ± 0°) were achieved by plasma treating the OTS-*t*-glass prism for different exposure times. The methodology for surface preparation (SI, Section 1) and resulting surface chemistry (evaluated by x-ray photoelectron spectroscopy and attenuated total internal infrared spectroscopy) has been described in great details elsewhere (3).

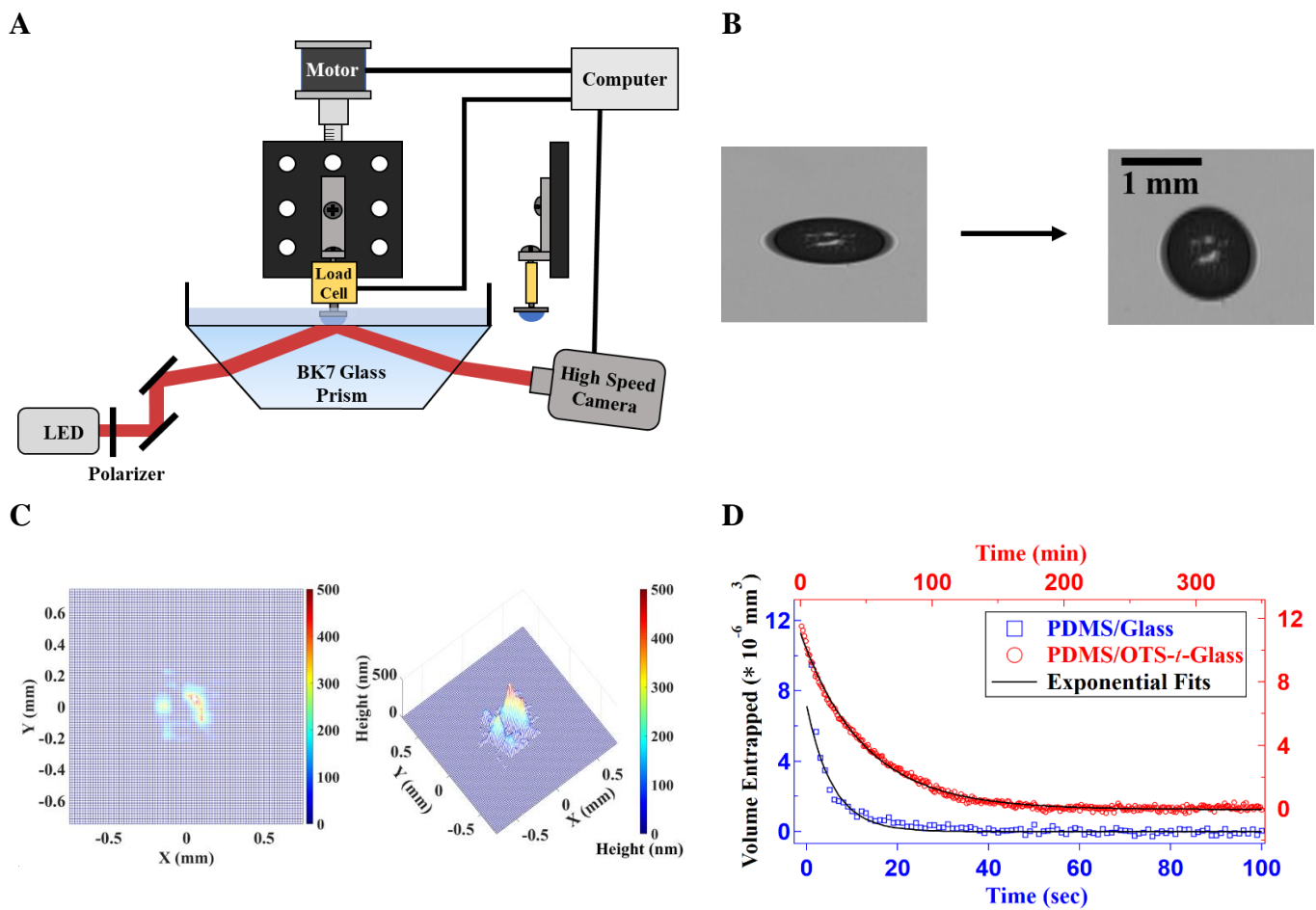

**Fig. 1.** (A) Schematic (not to scale) of the frustrated total internal reflection imaging setup. (B) FTIR image captured by the camera (left image) for a PDMS elastomer/hard substrate (51° water contact angle) contact underwater. Right image shows the ellipse backward transformed image using the ellipse's aspect ratio (C) Height map (top and side view) calculated by normalizing the gray-scale image in part b) (right side) with background and conversion to corresponding heights using reflectance equation (Methods). (D) Volume entrapped versus time data for PDMS/glass (empty blue squares) and PDMS/OTS-*t*-Glass contact (empty red circles) and the corresponding exponential fits (solid black lines) to extract the decay constant. Volumes are calculated by integrating the height maps every time step.

Fig. 1B shows the FTIR image captured (at t = 1.5 sec) for PDMS/intermediate wettability glass (with water contact angle of 51° ± 1.8°) contact underwater. The dark region represents the pixels for which the evanescent wave is totally frustrated due to contact of PDMS with the substrate, resulting in refraction of incident light into PDMS, whereas the white region represents pixels where the wave is partially frustrated due to presence of entrapped water. The captured image is elliptical due to optical light path and back transformed (Fig. 1B) right) using aspect ratio of ellipse (calculation in Methods) to perform analysis. A look-up table (SI, Fig. S2) between the modulus of total reflectivity squared (reflectance) versus film thickness was prepared (procedure in Methods) using the Stokes' relations and Fresnel equations for p-polarization, which was then used to convert the reflectance of every pixel to the height of water entrapped (27). Any pixel with water film thickness less than 3 nm is considered as contact due to limit in resolution for this technique. Fig. 1C shows the top and side view of the heightmap corresponding to the gray-scale image in Fig. 1B. The obtained spatiotemporal (height maps at each time step) data is used for calculating the volume of entrapped water by integrating the height maps at each timestep, represented in Fig. 1D.

**Results**

**Water Evacuation Dynamics.** Fig. 1D shows the water evacuation profiles for two extreme cases of wettability, i.e. PDMS/glass and PDMS/OTS-$t$-glass contact underwater. The volume of entrapped water was observed to decrease exponentially with time after initial contact irrespective of the substrate chemistry and the PDMS elastic modulus (G). The decay constants (K) for different cases were obtained by fitting the volume versus time data to equation, $V(t) = V_0 e^{-Kt}$, and the results are shown in Fig. 2. In Fig. 1D, it can be noted that the time required to evacuate water from a hydrophobic-hydrophobic entrapment is ~ 3 orders of magnitude higher than that required for the hydrophilic-hydrophobic entrapment. The results are surprising as we expected that the water evacuation would be faster from two hydrophobic surfaces entrapment due to their thermodynamic inclination to be in contact. We expect the contact between two hydrophobic surfaces underwater to be dry based on earlier measurements using surface-sensitive sum-frequency generation spectroscopy (SFG) and the measured work of adhesion (~ 80 mJ/m$^2$ same as thermodynamic prediction) between PDMS/OTS-$t$-glass underwater. In comparison, we expect the glass-hydrophobic contact should always contain a thin layer of water, which has also been verified by SFG (3).

**Effect of PDMS modulus and substrate wettability on rate constant.** In Fig. 2A, we plot the decay rates (K) for water evacuation as a function of bulk modulus (G) of PDMS for substrates with different wettability represented by water contact angles. The positive slope for all the trends represents faster evacuation for higher G. Further, the value of K increases with a decrease in substrate contact angle. It is evident from the graph that the effect of modulus (less than 1 order) is very small compared to that of surface wettability (more than 3 orders) for the parameter range explored here. Fig. 2B shows a plot of log(K) versus contact angles for PDMS lenses with different modulus. There is a notable, apparent change in slope around ~42° contact angle (intersection of the two linear fits) for all the cases of G, indicating towards a transition in the mode of evacuation dynamics. The transition is also clearly visible in a log(K) versus cosine of contact angle plot (SI, Fig. S6).

**Initial Contact Formation and Modes of Evacuation.** To shed more light on the changes in the evacuation rates Fig. 3 shows the spatiotemporal heightmaps for initial contact formation and the following water evacuation for interfaces between PDMS lens (0.7 MPa) and three different substrates (glass (0°), intermediate glass ($i$-glass) contact angle (35.9°± 1.0), OTS-$t$-glass (108° ± 3°)). Corresponding gray-scale images are shown under each heightmap. t = 0 is chosen arbitrarily from initial frames of each of the three cases where the elastohydrodynamic deformation can be noticed clearly.

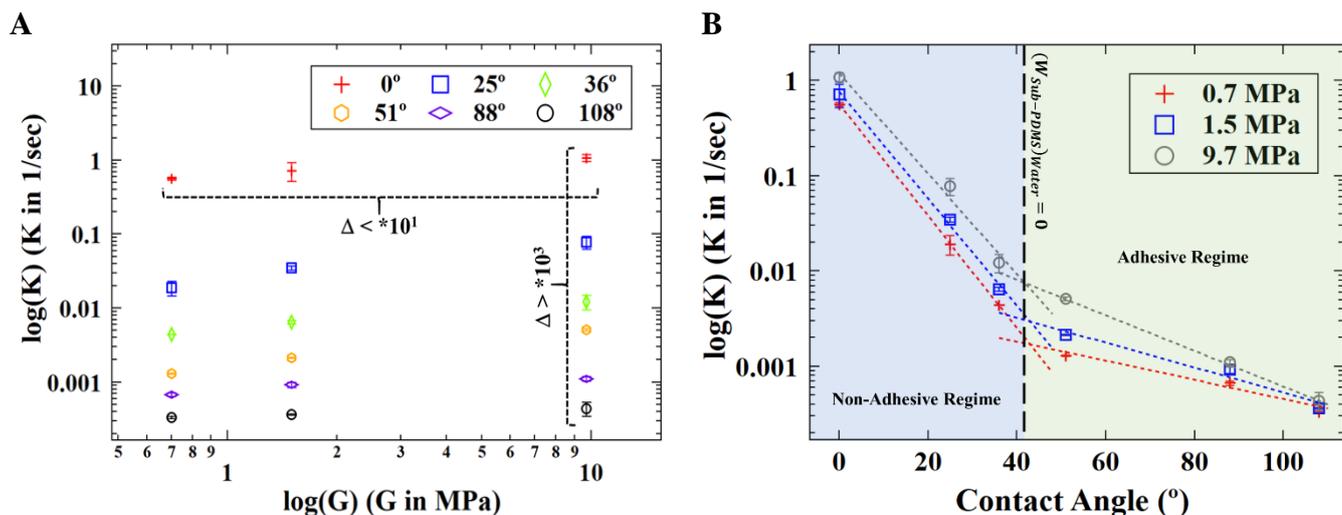

**Fig. 2.** (A) Dependence of water evacuation rate constant on elastomer modulus (G) for different substrate wettability represented by the log-log plot. The vertical and horizontal dashed black line segments demarcate the role of substrate chemistry and modulus by comparing the change in order of magnitude in K values. (B) Log(K) versus contact angle plots for three moduli of PDMS. The plots clearly point towards a transition in evacuation dynamics as the slope of data changes about ~42°, i.e. the point at which underwater thermodynamic work of adhesion is zero (dashed vertical black line). The dashed vertical black line separates the adhesive and non-adhesive regimes. Dashed colored lines are exponential fits to K versus contact angle data for two segments about the transition point for three moduli of lenses. All the rate constants are calculated from the volume entrapped versus time data as shown in Fig. 1D.

Fig. 3A shows the entrapment of water in bell shape (t = 0) and the subsequent growth of area (t = 0.14 s) outside the bell as the normal displacement progressed to reach to the final normal load. A continuous thin film of water is observed throughout the apparent contact area with the additional symmetric puddle (water bell) in central position (Fig. 3A and SI, Fig. S4). The puddle height decreases with an increase in puddle radius (t = 0.140 s to t = 3.490 s). Additionally, the 2D line heightmaps (SI, Fig. S4) for the contact formation shows the gradual height decrease of thin film and the puddle without any contact between the PDMS and glass suggesting that the water evacuates axisymmetrically without any barrier in flow path. Even though the final state (t = ∞) of the contact interface seems to be completely dry, certainty of intimate atomic contact cannot be guaranteed due to limitation in the film thickness resolution (~ 3 nm) of the FTIR technique. Previously, the SFG (3) results have shown that the contact interface between hydrophobic PDMS and hydrophilic sapphire surfaces are not dry. Molecular dynamics simulations have also proved that even a highly hydrophobic surface is fully hydrated if the second surface is highly polar (28).

Fig. 3C left-hand panel shows the initial (t = 0) water entrapment for the PDMS lens approaching OTS-*t*-glass underwater. Suddenly after the small puddle formation, the puddle gets asymmetrically deformed (as evident in t = 0.014 s). In fact, it is apparent from 2D cross-sections of Fig. 3C (line heightmaps, SI, Fig. S4) that the symmetry of the deformed lens is broken (instability) when the peripheral gap thickness is still ~100 nm (for the case with softest lens) and, immediately after, a part of the peripheral bell appears to snap into contact. This leads to enhanced asymmetry (SI, Fig. S4) and breakdown of the puddle into multiple smaller puddles (t = 0.107 s). Further, it is evident from the 2D line heightmaps (SI, Fig. S4) that majority of the apparent area is true contact (in FTIR limit) between the PDMS and OTS-*t*-glass at the motor stop time. Fig. 3C right-hand panel illustrates the water evacuation from the small puddles entrapped between PDMS lens and OTS-*t*-glass contact where the height of individual puddles goes down with a decrease in puddle radius, in contrast with that of the

PDMS/glass contact. In this case, no continuous thin film for the apparent contact is observed suggesting barriers in the water flow path.

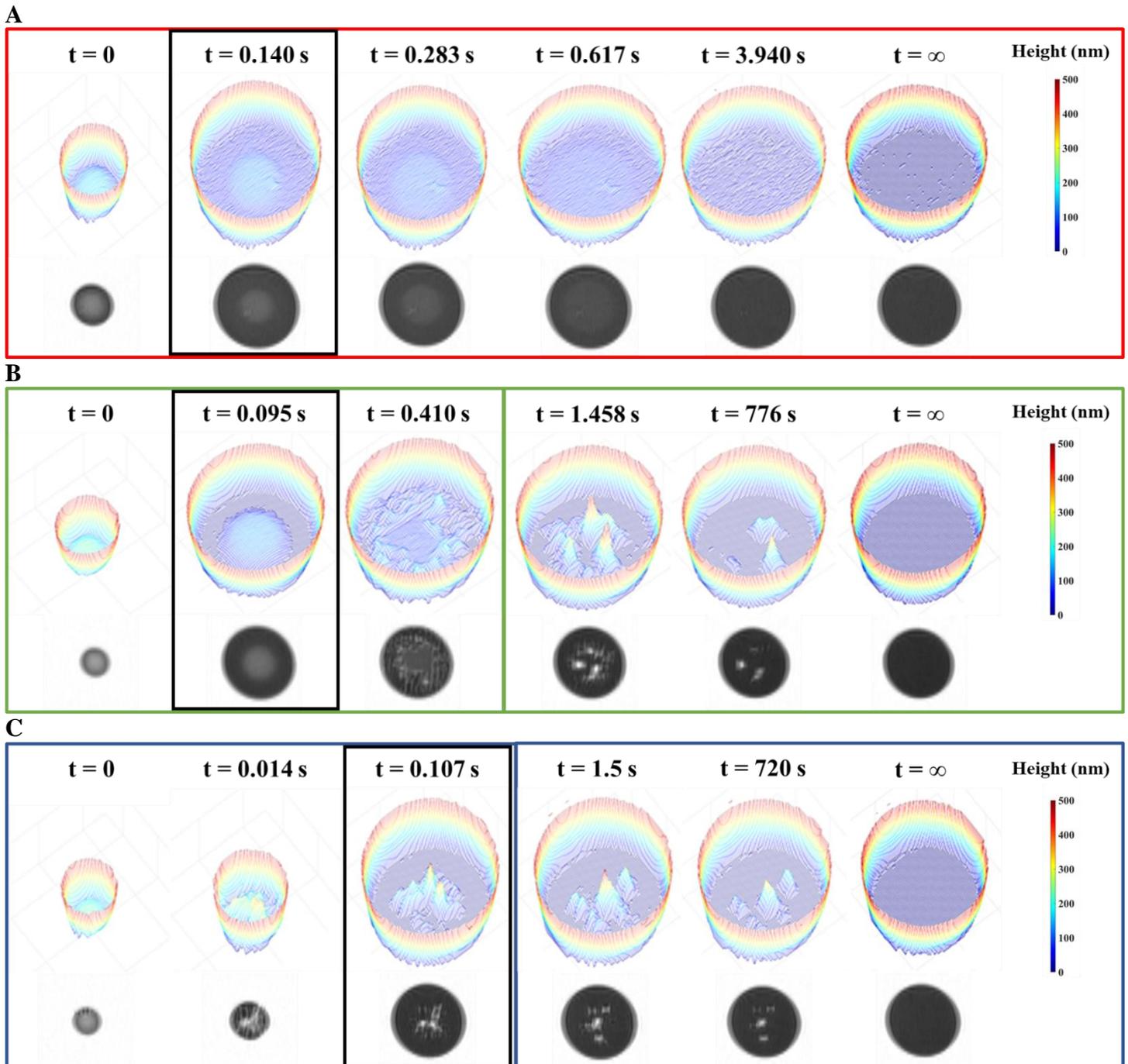

**Fig. 3.** Spatiotemporal height maps and corresponding gray-scale images for PDMS/glass (red panel) (A), PDMS/intermediate contact angle (35.9°± 1.0) glass (green panel) (B) and PDMS/OTS-*t*-glass (blue panel) (C) interface. For all the cases, black outline enclosed box represents the frame where motor stopped. The left and right panel for the case (B) and (C) represent frames of two separate repeats performed at 1000 frames/sec and 24 frames/sec to capture the initial short-time deformation and contact formation and long-time water evacuation from the entrapments. For (A), all the frames belong to a single experiment performed at 24 frames/sec. The data presented is for the case with PDMS lens of 0.7 MPa (G) and the observed trends are invariant to G.

Fig. 3B shows the initial contact formation (left-hand panel) and the water evacuation (right-hand panel) from the PDMS lens/*i*-glass interface. The initial behavior of the system is similar to the PDMS/glass interface where first we observe water entrapment (t = 0) in a bell-shape due to elastohydrodynamic deformation followed by a decrease in puddle height and an increase in puddle radius (t = 0.095 s). With the reduction in puddle height, we start to observe an instability and waviness in the water film thickness with water entrapped in a combination of puddles and thin film at the center region (t = 0.410 s). We then observe that the entrapped water in combined (puddle-film) state (t = 0.410 s) further coalescing with existing puddles growing and nearby film disappearing (t = 1.458 s). The behavior suggests that the increased elastic energy from thickening individual puddles is balanced by decreasing surface energy of growing dry contact. Similar behavior of increment in the height of puddles is also observed in the PDMS lens/OTS-*t*-glass interface case but the process there is much faster and completed even before the motor stop. The evacuation from the smaller puddles at PDMS lens/*i*-glass interface optically followed the mechanism of that for PDMS lens/OTS-*t*-glass interface where the heights of the puddles are found to decrease with time without presence of a continuous thin film. Despite the similarity in the evacuation mechanism, the dynamics of evacuation from PDMS lens/*i*-glass interface is much faster than that of PDMS lens/OTS-*t*-glass interface (Fig. 2 and Fig. 3), which further emphasizes the role of adhesion in evacuation mechanism.

**Human Thumb Contacts.** To explore the generality of the observations to realistic contacts, we study the dynamics of biological underwater contacts of human digit (thumb outermost layer: stratum corneum) with hard hydrophilic (bare glass) and hydrophobic substrates (OTS-*t*-glass) in Fig. 4 (larger area with corresponding gray-scale images in Fig. S5, SI). For both the contacts, it is observed (Fig. S5, SI) that the protruding ridges of the thumb tend to come in contact (t = 0) with the hard substrate in an unconnected fashion. Further, the contact region evolves underwater in two-step coalescence process (from t = 0 to t = 0.04 s and t = 0.04 onwards) similar to that reported earlier for the dry human figure-tip/glass contact (29). The number of junction points increase (t = 0 to t = 0.04 s) followed by the growth in their areas (t = 0.04 onwards) resulting into the connection of those junction points into continuous ridges. The growth is usually fast (within few seconds) and ascribed to the plasticization (reducing the elastic modulus to few MPa from ~ 1 GPa) of stratum corneum due to hydration (30, 31).

Fig. 4A shows the evolution of water heights in a randomly selected region (for evolution of complete region see Fig. S5, SI) of thumb/glass and thumb/OTS-*t*-glass contact. It is clear from these spatiotemporal heightmaps that the thumb valleys act as channels for collection of water from the ridges (t = 0.10 s), facilitating the growth in contact area. After ridges connect (t = 0.40 s), for the case of thumb/glass interface (Fig. 4A, top panel), we observe that the water valleys are continuous and gradually narrowed with time, enhancing the width of ridges and, hence, the contact area. Eventually some of the ridges joined due to the diminishing valleys for some regions of the thumb (t = 10 s), suggesting evacuation of water through valleys to large extent. In the case of thumb/OTS-*t*-glass interface (Fig. 4A, bottom panel), we observe an instability (t = 5 s) in the valley region following the ridges connection (t = 0.40 s) and subsequent water evacuation. There is a sudden breakdown in the continuity of flow path of the valleys resulting into unconnected entrapments of water in valleys (t = 5 s), causing evacuation of water to lesser extent. The long-time dynamics shows that the evacuation from these entrapments happen as discreet puddle-wise events (Video S8, SI). To quantify the dynamics of water evacuation, we plot average volume versus time curves for the thumb/glass and thumb/OTS-*t*-glass contact in Fig. 4B. Initially the water evacuation for both the cases follow similar dynamics with deviations starting at ~ 7 sec. Interestingly, the asymptotic volume, after the evacuation process, is much higher for the case of thumb/OTS-*t*-glass contact.

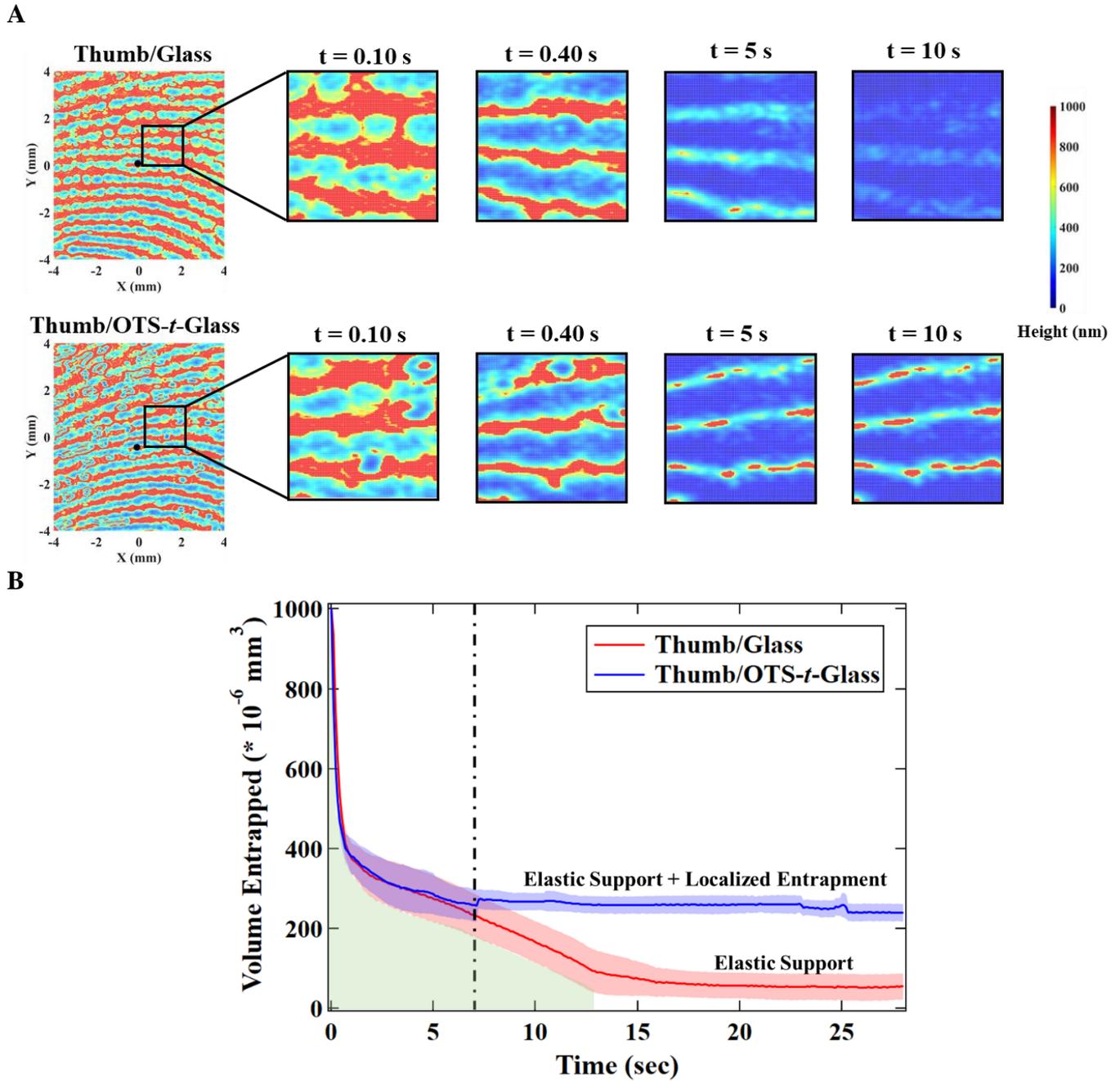

**Fig. 4.** Spatiotemporal height maps for water in a localized region (marked by black solid box) for a human thumb in contact with glass (top panel) and OTS-*t*-glass (bottom panel) (A). Average (15 non-overlapping regions) volume evolution for thumb/glass (solid red line) and thumb/OTS-*t*-glass (solid blue line) contact with corresponding standard deviations (light red and blue shaded area) (B) highlighting similarity in evacuation dynamics till ~ 7 sec (marked by vertical dashed-dotted black line) and differences in asymptotic volume for the two cases. The light green shaded region represents the water evacuation through valleys for glass case till ~ 13 sec and for OTS-*t*-glass case till ~ 7 sec. The evacuation limit for glass is reached due to elastic support of stratum corneum whereas that for OTS-*t*-glass is reached due to elastic support along with localized entrapment of water puddles shown in part A (bottom panel).

## Discussion

Here, we seek to understand the trends in dynamics of water evacuation (Fig. 2) and provide deeper insights into the process of contact formation, subsequent evacuation (Fig. 3 and Fig. 4) and their links with the observed rate constants (K). Fig. 2 demarcates the role of PDMS modulus and surface wettability in the observed evacuation dynamics. The faster dynamics for higher G could be due to lower apparent contact area at same normal force, hence, a smaller water evacuation path. Another possibility is that the lenses with higher G cannot conform to the surface roughness which can result in wider channels for water evacuation. The significant effect of surface wettability on evacuation dynamics and the change in slope in log(K) versus water contact angle (θ) around ~42° has been rationalized using a thermodynamic argument below.

**Transition in Evacuation Dynamics.** The crossover in slopes (Fig. 2B) of rate constants is intriguing as it points towards a transition in the evacuation mechanism. The significance of the transition point with respect to substrate wettability can be related to the corresponding adhesion energies for the system using the Young (32) and Dupré (33) equations and available experimental values.

The Dupré equation for two continuum media (*Sub1* and *Sub2*) interacting in a third medium (say underwater) can be written as Eq. 1.

$$(W_{Sub1-Sub2})_{Water} = \gamma_{Sub1-Water} + \gamma_{Sub2-Water} - \gamma_{Sub1-Sub2} \tag{1}$$

where $(W_{Sub1-Sub2})_{Water}$ is the thermodynamic work of adhesion underwater and $\gamma'$s are interfacial energies between *Sub1*, *Sub2* and *water*. Subtracting from equation 1 the Dupré expression for *Sub1-Sub2* in air (third medium), and substituting the difference in interfacial energies underwater and in dry case for *Sub1* and *Sub2* in terms of water contact angles ($(\theta_{Sub1-water})_{air}$ and $(\theta_{Sub2-water})_{air}$) and water interfacial energy ($\gamma_{Water}$) using the Young's equation we get, Eq. 2.

$$(W_{Sub1-Sub2})_{Water} = (W_{Sub1-Sub2})_{air} - \gamma_{Water}[Cos(\theta_{Sub1-water})_{air} + Cos(\theta_{Sub2-water})_{air}] \tag{2}$$

For this study, *Sub1* is hard substrate (*Sub*) and *Sub2* is PDMS. Plugging in the interfacial energy for water-air interface (72.8 mJ/m$^2$) and the contact angle (110°) of water on PDMS in Eq. 2, we get Eq. 3.

$$(W_{Sub-PDMS})_{Water} = (W_{Sub-PDMS})_{Air} - \gamma_{Water} * Cos\theta_{Sub-Water} + 25 \tag{3}$$

Assuming that the observed transition point (~42°) represents the point where $(W_{Sub-PDMS})_{Water}$ is equal to zero, we estimate the $(W_{Sub-PDMS})_{Air}$ to be equal to 29 mJ/m$^2$. The estimated dry work of adhesion for ~42° water contact angle is in agreement with the experimental value (30 ± 3 mJ/m$^2$) measured during the loading cycle using the JKR method (3). The change in slope where the underwater thermodynamic work of adhesion is zero suggests that formation of dry contacts or barriers in water flow path is responsible for slowing down the evacuation rates, as seen in Fig. 1D and Fig. 2. The connected barriers in flow path presents the argument of sealing mechanism or water entrapment in smaller puddles.

We highlight that the observed point (~42°) of transition in dynamics should not be considered as special number. The transition point depends upon the thermodynamics of the three phases. Any change in the surface energy of the elastomer or selected liquid will shift the transition point to either lower or higher side. The open question is to synthesize the elastomers with varying surface energies and hence design a universal evacuation curve. Further, modelling the dynamics about the transition point requires a broad consideration of parameter space which is yet to be discovered, especially, the influence of roughness seeks our attention and will thrive the field of underwater contact mechanics.

**Contact formation and subsequent evacuation.** Uniting our observed trend (Fig. 1D, Fig. 2 and Fig. 3) with the established knowledge of thermodynamics and material deformation allows us to explain the contact formation in terms of two sequential processes. First, for all the cases presented in Fig. 3, the PDMS lenses undergo elastohydrodynamic deformation due to liquid viscous forces which results in entrapment of water (confirmed from the height maps before contact, SI, Fig. S4). The contact region has water trapped in the center as a bell shape cavity with contact initiating at the periphery. The second process differs depending on the surface energy of the substrate. For the case of PDMS/glass contact, the water evacuates rapidly and smoothly through a thin film of water reducing its height with time (Fig. 3A). In the case of PDMS/OTS-*t*-glass contact, the puddle quickly and asymmetrically breaks into smaller puddles, which drain over a much longer period (Fig. 3C), limited by a tighter seal of the adhesive contact at the periphery.

Now, we discuss the similarities in the water evacuation dynamics for model system (PDMS lens/hard contact) with a complex reality (human thumb/hard contact). The water evacuation dynamics for both thumb/glass and thumb/OTS-*t*-glass contact is identical till ~ 7 sec (Fig. 4B) and can be ascribed to the presence of valleys which facilitate the drainage (Fig. 4A and Fig. S5). We observe striking differences between the dynamics of thumb/glass and thumb/OTS-*t*-glass contact after ~ 7 sec that can be explained with our knowledge of evacuation process for the model systems. The cleaned surface of stratum corneum is known to be highly hydrophobic (contact angle of 125° – 130°) (34) due to which the thumb/OTS-*t*-glass interface should behave in a similar fashion as PDMS lens/OTS-*t*-glass interface at microscopic scale. For initial region (till ~ 7 sec) the elastic modulus is high enough such that the adhesive forces have no effect on the dynamics. Subsequently, we observe sudden asymmetric break down of the valleys (Fig. 4A, bottom panel) for the case of thumb/OTS-*t*-glass interface which results in localized water entrapments. The instabilities in the valleys for thumb/OTS-*t*-glass contact seem similar to the breakdown of a large puddle into multiple smaller puddles for the case of PDMS lens/OTS-*t*-glass contact. Additionally, we observe asymptotic volume for both thumb/glass and thumb/OTS-*t*-glass contact. This volume represents static condition in the observation timescale wherein the elastic stiffness of the ridges in the fingerprint support cavities of water in the valleys between. The previously reported drop in stiffness of the stratum corneum of a human finger with hydration/moisturization is likely changing this elastically supported volume as the contact evolves. The larger value of static volume of trapped water associated with thumb/OTS-*t*-glass contact is consistent with the notion that sealed puddles of incompressible water are contributing to the resistance of the fingerprint ridges to further collapse.

Here, we would like to point out that for the hydrophobic-hydrophobic (PDMS lens/OTS-*t*-glass) contact formation under water (Fig. 3C), we have made new observations which are discussed below.

**Water puddle breaks down before contact, an instability for hydrophobic-hydrophobic entrapment.** The instability in the deformation of a single puddle for the case of PDMS lens approaching OTS-*t*-glass before any direct contact (Fig. 3C and SI, Fig. S4B) points toward the presence of long-range hydrophobic forces, popularly known as solvation forces. When two hydrophobic surfaces approach each other underwater, a water-vapor cavity formation takes place that applies negative Laplace pressure on both the surfaces resulting into the observed attraction between them (35). The cavity formed is a transient state which is lower in energy than the two hydrophobic surfaces separated by some water layer but is still higher in energy when compared to the final thermodynamic equilibrium state of direct solid contact (35,

36). These hydrophobic solvation forces cannot be accounted for by the Derjaguin-Landau-Verwey-Overbeek (DLVO) or Lifshitz theory and has now been proven to exist, experimentally as well as through simulations and theory (28, 35, 36). Using surface force apparatus, it has been shown that these forces start to show up at separations of 20-100 nm (depending upon the hydrophobicity) for two rigid hydrophobic surfaces underwater (35, 36). We also observe that during instability only a part of periphery touchdowns the substrate (SI, Fig. S4B). We speculate that as the water evacuation rates are high initially due to the high stored elastic energy, hence hydrostatic pressure, the peripheral edge opposite to the edge in contact is found to stay out of contact for some time to facilitate water evacuation. After a certain time, the entire peripheral area is found to seal with the smaller water puddles trapped in between. The evacuation mechanism for these smaller puddles following the sealing is unobservable and has been hypothesized below.

**How is water escaping from sealed hydrophobic-hydrophobic entrapment?** The water evacuation from the smaller puddles entrapped between the PDMS lens and OTS-*t*-glass (Fig. 3C) can happen in multiple ways. First, it is possible that the PDMS elastomer, even though known to have no swelling in water, is absorbing some water. However, all PDMS lenses are kept underwater for ~6 hr prior to the data collection and no differences in the decay constant (K) is observed even if the presoaking time is increased. Second, a driving pressure difference between the entrapped water and the ambient conditions exist which can play a role in diffusion through the PDMS material. The differences in the rate constants for evacuation (K values reported in Fig. 2) for the glass substrate with 88° and 108° contact angle do not support the diffusion argument as both of the glass substrates make adhesive contact with PDMS underwater and the K values should be similar if the mode of evacuation is diffusion through the PDMS material. Lastly, it is possible that the water is evacuating out along the interface through gaps smaller than the measurable length scale. It is well known that even though all the glass surfaces used are optically smooth, these surfaces are rough at nm length scale (3). The roughness at smaller length scales can remain as an interconnected channel facilitating the water evacuation along the interface (15). Further, the final equilibrium state can have majorly dry regions with water sealed at small scale roughness. The effect of surface roughness on water evacuation is difficult to investigate using the FTIR technique as the rougher surfaces will contribute to the light scattering and make the data analysis impractical (27). On the other hand, making rough or patterned PDMS structures are totally feasible and will be the scope of our future studies.

**Conclusion**

Competition between adhesive forces and fluid drainage can play a crucial role in collisions between deformable objects (37). While adhesion energetically favors contact between hydrophobic contacting pairs, that very affinity dynamically leads to entrapment of isolated puddles, thus delaying their ultimate contact. The paradoxical relationship between kinetics and thermodynamics may explain why underwater adhesive designs which rely on surface hydrophobicity alone do not produce the best performance. Patterned or special morphology is necessary to increase evacuation rates (12).

We have demonstrated that increasing adhesive interaction between soft bodies leads to as many as 3 orders of magnitude in dynamics inhibition of their contact formation due to tighter sealing of trapped water, in a controlled system with simple geometry. In the complex practically important system – human thumb contact underwater – we found that competition between the same physical drivers *plus surface texture* lead to a logically consistent, non-trivial result: the surface texture facilitates the initial drainage till the point where the adhesive forces overcome the elastic response of the thumb, that prevents longer-term approach to more fully intimate contact due to better sealing of puddles trapped between the adhesive contact.

Evolution of underwater contact area is important as it determines the creation of friction which is essential in everyday life. It would have been impossible to even hold a wet glass of water without any intimate contact. Though the increase in real contact area at microscopic level to high tangential frictional force in dry haptics has been well studied, the augmentation of human tactile sensing underwater is challenging as the primary factors include the length scale and time scale of contact (29, 38). Quantifying the area growth laws and exploring the wet haptics will be scope of our future studies.

## Methods

### In-Situ Contact Experiments and Analysis.
**FTIR setup and experiments.** Schematic of the FTIR setup is shown in Fig. 1A. The top surface of the dove prism is illuminated by a polarized mounted LED (M660L4, λ ~ 665 nm, bandwidth 20 nm, obtained from Thorlabs). The dove prism glass material is transparent to the light wavelength and the prism surface is optically flat. Incident angle ($\theta_1 = 64.4° \pm 0.1°$) with respect to glass surface normal of the p-polarized light (λ = 665 nm) has been selected between the critical angle for glass/water and glass/PDMS interface. At FTIR condition, an evanescent wave propagates along the interface with its intensity decaying as an exponential function along the Z direction in the second medium. When a third medium comes into evanescent wave, a portion of the energy is refracted in third medium. This energy loss carries the information of film thickness of the second medium. Initial quantitative model (39, 40) for film thickness assumed a simple truncation of evanescent wave intensity whereas later (27) it was proved that multiple transmissions and reflections, and also the light polarization plays a crucial role in accurate predictions of film thickness. Thus, a polarizer was used in the light path to set the p-polarization of the incident light. When a PDMS lens approaches the glass prism underwater, the FTIR phenomenon starts to occur at a separation comparable to the wavelength of light. The Z-stage is connected to PDMS lens through a load cell to control the vertical displacement of the lens to achieve an initial load of (7 ± 0.2) g after which the motor is stopped.

**Data Acquisition and Image processing.** The reflected intensity profiles from the internal face of the dove prism were collected using a high-speed camera (VEO410 procured from Phantom) with a video resolution of 128×128 pixels$^2$ at 24 and 1000 fps. The fast frame rate (1000 fps) was good enough to visualize the initial contact formation whereas the slow frame rate (24 fps) allowed us to capture long time videos for studying water evacuation dynamics from the hydrophobic-hydrophobic entrapment. Fig. 1B (left image) shows a raw grayscale frame of a video which shows both, the region of dry contact or dark pixels, i.e. PDMS in contact with the hard substrate, as well as the bright pixels which corresponds to the case of thin water film in between the PDMS and substrate. Before performing the experiments, a water droplet is put on the top of a dry hydrophobic prism which gives a perfectly circular boundary line for the droplet and the image is captured. The captured image is elliptical and used for calculating the aspect ratio for back transformation of the acquired data and the incident angle. A transformed frame of a video was then divided by the background image (frame with total internal reflection of the light at every pixel or t = 0 image) of that video. The normalized intensity at each pixel (reflectance) was then converted into corresponding water film thicknesses ($h$) using a Look-Up Table for p-polarization presented in Fig. S2 to render a 3D height map. The Look-Up was created using the relationship between normalized reflection intensity (reflectance, $|\hat{r}|^2$) and the film thickness. This reflectance equation (Eq. 6) accounts for multiple reflections and transmission summed (geometric series) using Stokes' relations and for light polarization through Fresnel equation, and has been reported in great details earlier (27, 41). The lateral resolution of the camera is ~ 15 μm/pixel. To verify the FTIR technique and our analysis protocol, we measure (SI, Fig. S3) the air film thickness between a convex glass lens (radius of curvature: 386 mm, obtained for Thorlabs) and glass prisms (pristine and OTS coated) under a 20g load. The measured gap thicknesses showed an excellent agreement with the Hertzian contact theory.

$$\hat{r} = \frac{\left(\frac{n_2\cos\theta_1 - n_1\cos\theta_2}{n_2\cos\theta_1 + n_1\cos\theta_2}\right) + \left(\frac{n_3\cos\theta_2 - n_2\cos\theta_3}{n_2\cos\theta_3 + n_3\cos\theta_2}\right)\exp\left(j\left(\frac{4\pi h}{\lambda}\sqrt{n_2^2 - n_1^2\sin^2\theta_1}\right)\right)}{1 + \left(\frac{n_2\cos\theta_1 - n_1\cos\theta_2}{n_2\cos\theta_1 + n_1\cos\theta_2}\right)\left(\frac{n_3\cos\theta_2 - n_2\cos\theta_3}{n_2\cos\theta_3 + n_3\cos\theta_2}\right)\exp\left(j\left(\frac{4\pi h}{\lambda}\sqrt{n_2^2 - n_1^2\sin^2\theta_1}\right)\right)} \quad (6)$$

The total reflected amplitude ($\hat{r}$) is related to indices of refraction of glass ($n_1$), water ($n_2$) and PDMS ($n_3$). $\theta_1$, $\theta_2$, and $\theta_3$ denotes the incident angles for glass/water, water/PDMS and PDMS/water interface. $j$ is an imaginary number and $h$ is the gap or water film thickness between prism and PDMS lens.

**Human thumb Experiments.** The right thumb of the subject (one of the authors, male, 25 y old, right-handed Asian) was brought into contact at an angle of 0° with glass prism. The glass was loaded to a constant force of 10 ± 0.5 N (force sensor installed at the bottom of the glass prism in Fig. 1) and was held in position for 1 minute. Before the experiment, the thumb of the subject was first washed with commercial soap, then wiped with ethanol using a nonwoven paper and finally rinsed with ultrapure water (Millipore filtration system, volume resistivity of 18.2 MΩ.cm) for 10 seconds.

**Acknowledgements and Funding Sources**

We thank Professor Joelle Frechette, Professor Andreas Carlson and Sukhmanjot Kaur for helpful comments. We acknowledge the financial support from the National Science Foundation (NSF) (partly by NSF DMR 1610483, and NSF DMR 1508440). N.K. and A.D. were supported by the National Science Foundation. H.K. acknowledges the University of Akron for research start-up support.

**Author Contributions**

N.K., M.S., H.K. and A.D. designed research; M.S. and N.K. performed research; M.S., N.K., H.K. and A.D. analyzed data; N.K. wrote the manuscript; M.S., H.K. and A.D. reviewed and edited the manuscript.

[1]M.S. and N.K. contributed equally to the work.

# Supplemental Information for "Adhesive forces inhibit underwater contact formation for a soft-hard collision"

Mengyue Sun[1], Nityanshu Kumar[1], Ali Dhinojwala[*], Hunter King[*]

*The University of Akron, Department of Polymer Science, Akron, OH - 44325*

[1]M.S. and N.K. contributed equally to the work.
[*] Correspondence should be addressed to ali4@uakron.edu and hking@uakron.edu

## SI, Section 1. Materials Synthesis and Preparation

**Synthesis of PDMS lenses.** Soft and smooth elastomeric lenses of varying modulus were synthesized by cross-linking the PDMS. A desired range of moduli was achieved by changing the cross-link molecular weight ($M_c$) of the monomers in accordance with the network theory which states that the elastic modulus of a perfectly cross-linked system is inversely proportional to the $M_c$ and can be estimated using relation, $E \cong \rho RT/M_c$, where $\rho$ is the density of the polymer, R is the gas constant, T is absolute temperature, and Mc is the cross-link molecular weight. Vinyl-terminated PDMS of three different molecular weights ($M_w$) (DMS V-05 [Mw = 800 g/mol], V-31 [Mw = 28,000 g/mol] and V-41 [Mw = 62,700 g/mol]) as monomers, tetrakis-dimethylsiloxysilane as tetra-functional cross-linker and platinum carbonyl cyclo-vinyl methyl siloxane as catalyst were procured from Gelest Inc. Monomer and cross-linker were first mixed in a molar ratio of 4.4 in an aluminum pan, then the catalyst was added as 0.1 wt% of the total mixture and finally the batch was transferred to a syringe with a needle for casting. Before casting the lenses, the syringe was place in a vacuum chamber for 5 mins to remove all the air bubbles entrapped while physical mixing. As the process of mixing and vacuum application for air bubble removal is prone to cross-linker evaporation, the initial ratio of monomer:cross-linker was set higher than the required stoichiometric ratio for reaction, i.e. 4. Previously, this methodology has proved to avoid the cross-linker evaporation below stoichiometric ratio and, hence, minimize the adhesion hysteresis due to viscose-dissipation of the unreacted groups in the network (1, 2). Lenses for JKR measurements and FTIR experiments were casted on fluorinated glass petri-dishes. The lenses were casted on a fluorinated dish as it has surface energy lower than PDMS, hence, the PDMS mixture drop maintains a hemispherical shape. The dishes were sealed with glass led on top and were cured at 60° C for 3 d. Further, the lenses were transferred to cellulose extraction thimble for Soxhlet extraction where toluene refluxes at 130° C for 48 h. PDMS lenses were again transferred to a fluorinated dish and dried in air for 12 h. Finally, the lenses were vacuum dried at 120° C for 16 h and then used for experiments. The radius of curvature was measured by fitting a 3-point circle to the image obtained using an optical microscope (Olympus). Lenses with curvatures below the capillary length of PDMS were selected for experiments (1.25 ± 0.05 mm for modulus measurements through JKR framework and 1.50 ± 0.05 mm for FTIR measurements) to ensure perfect hemispherical shape. Larger lens size was used for FTIR experiments to have large contact area for visualization through setup. The PDMS lenses of 0.7, 1.5 and 9.7 MPa elastic moduli were obtained (SI, Section 2) in the process. The clean base-bath treated borosilicate glass petri dishes were immersed in a 0.1 wt% solution (nitrogen purged throughout) of heptadecafluoro 1,1,2,2 tetrahydrodecatrichloro silane (procured from Gelest Inc.) (in toluene) for 8 h to deposit the fluorinated monolayer. After solution treatment, the petri dishes were thoroughly rinse with toluene, ethanol and deionized water and then dried in oven.

**Preparation of varying wettability glass prisms.** Uncoated glass dove prisms (N-BK7) for FTIR experiments were procured from Edmund Optics. The received prisms were sonicated in toluene, acetone, ethanol and then deionized water for 1 h each. Prisms were blown dry with nitrogen and then plasma sterilized for 5 min to remove any sort of contaminants. One of these prisms was kept as control glass. Another cleaned prism was dipped into a 0.1 wt% OTS (procured from Gelest Inc.) (in toluene) solution for 8 h and the nitrogen was purged into the solution. The OTS-$t$-glass prism was then rinsed with copious amount of toluene to get rid of any untethered residues of OTS. The OTS-$t$-glass prism was then annealed at 120° C for 6 h to enhance ordering of hydrophobic monolayer and minimize the contact angle hysteresis. Three measurements at different spots yielded a static contact angle of 108° ± 3° on the OTS-$t$-glass prism. To obtain prism surfaces with different wettability, the OTS-$t$-glass prism was air plasma treated for 1, 2, 3, 4 and 10 s. The contact angles obtained on those surfaces were 87.7°± 0.3°, 51.2°± 1.8°, 35.9°± 1.0°, 25.4° ± 0.8°, 0° ± 0°. The long time (10 s) plasma treated prism (0° ± 0°) and the control glass prism showed same results for evacuation dynamics experiments. The OTS monolayer was also prepared on the silicon wafers (procured from Silicon Inc.) following the procedure reported in literature (2). The OTS treated silicon wafers were used for JKR contact mechanics experiments (SI, Section 2).

**SI, Section 2. In-situ Johnson-Kendall-Roberts contact mechanics experiments**

We performed JKR contact mechanics experiments (3) to extract the moduli of the synthesized PDMS lenses. The OTS silanized silicon wafer was loaded with a PDMS lens upto a maximum force of ~ 1mN and then retracted. During the process in-situ measurements of the contact radius ($a$) and normal force ($N$) was done. The experimental setup for these experiments has been shown in the SI appendix of a previous literature (2). Data has been plotted as $a^3$ versus $N$ in Fig. S1 and fitted to Eq. S1 in order to extract $E^*$ (effective modulus) and $W$ (energy of adhesion). $E^*$ depends upon the elastic modulus of the two materials and their Poisson ratio as ($1/E^* = (1 - v_{\text{sphere}}^2)/E_{\text{sphere}} + (1 - v_{\text{substrate}}^2)/E_{\text{substrate}}$). As the glass prism (substrate) is hard, the quantity $E_{\text{substrate}}$ tends to $\infty$. Using the Poisson's ratio for PDMS equals to 0.5 (for perfect elastic systems), we extract the elastic modulus for PDMS ($E_{\text{sphere}}$ or G as referred in main text).

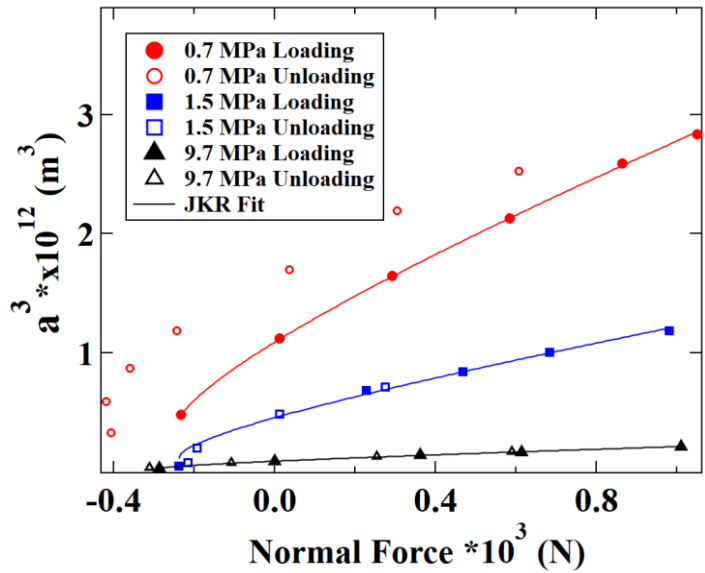

**Fig. S1.** Contact radius of PDMS lens with OTS substrate, a, and applied normal force was measured simultaneously in loading (filled symbols) as well as unloading (hollow symbols) cycle for three moduli of lenses 0.7 (red circles), 1.5 (blue squares) and 9.7 (black triangle) MPa. The normal load versus $a^3$ data is fitted (solid lines) to Eq. S1 (JKR equation) to obtain the work of adhesion and the modulus presented in the graph legend.

$$a^3 = \frac{3R}{4E^*}\left[P + 3\pi RW + \sqrt{6\pi RGN + (3\pi RW)^2}\right]$$
(S1)

**Table S1.** Elastomer modulus (G) and work of adhesion numbers calculated using Eq. S1 for PDMS approaching OTS coated substrate.

| $M_c$ (gm/mol) | $(W_{OTS-PDMS})_{Air}$ (approach) (mJ/m²) | Elastic Modulus (MPa) | $(W_{OTS-PDMS})_{Air}$ (retraction) (mJ/m²) |
|---|---|---|---|
| 800 | 50.4 ± 1.2 | 9.703 ± 0.001 | 55.4 ± 0.6 |
| 28000 | 42.2 ± 2.3 | 1.503 ± 0.032 | 42.4 ± 3.0 |
| 62700 | 44.2 ± 1.0 | 0.706 ± 0.012 | 69.2 ± 0.6 |

# SI, Section 3. Look-Up table for reflectance to height conversion and comparison of air film thicknesses measured experimentally with that calculated using Hertzian theory.

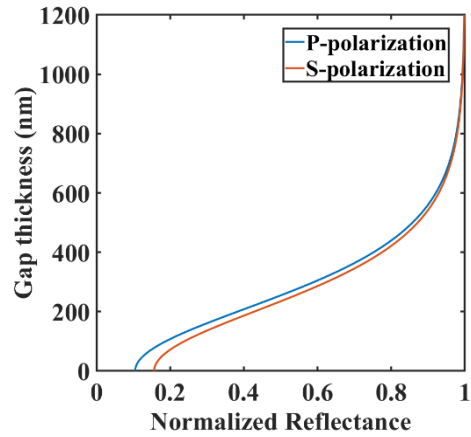

**Fig. S2.** Normalized reflectance as a function of gap thickness, also called Look-Up Table, used for conversion of reflectance to water film thickness. The parameters for Fresnel factors are, $n_1 = 1.52$ (glass), $n_2 = 1.33$ (water), $n_3 = 1.43$ (PDMS) and angle of incidence is $64.4° \pm 0.1°$.

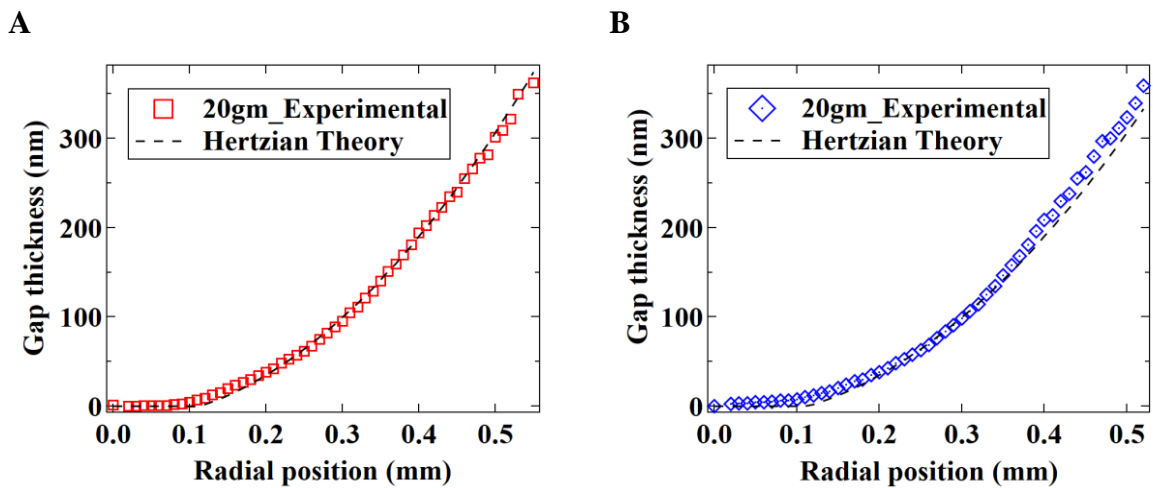

**Fig. S3.** Measured air film thickness for a glass lens/glass prism (hollow red square) (A) and glass lens/OTS-*t*-glass prism (hollow blue diamond) (B) contact for an external applied load of 20 gm. The corresponding predictions by Hertzian theory are represented by the dashed black line in each graph. The radius of curvature for the glass lens is 386 mm.

**SI, Section 4. 2D line heightmaps highlighting the water evacuation dynamics before motor-stop time for PDMS/glass and PDMS/OTS-*t*-glass interface.**

A

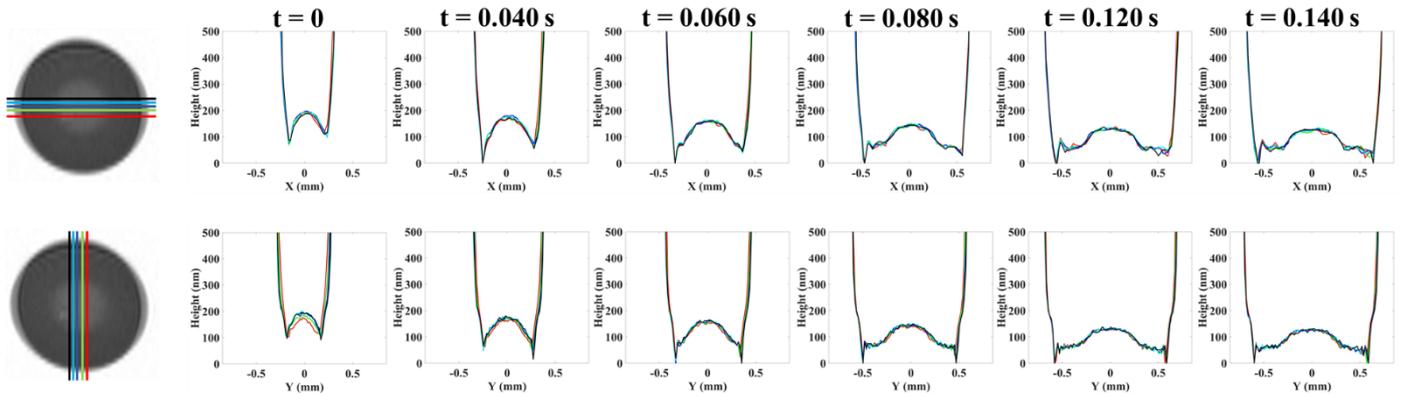

B

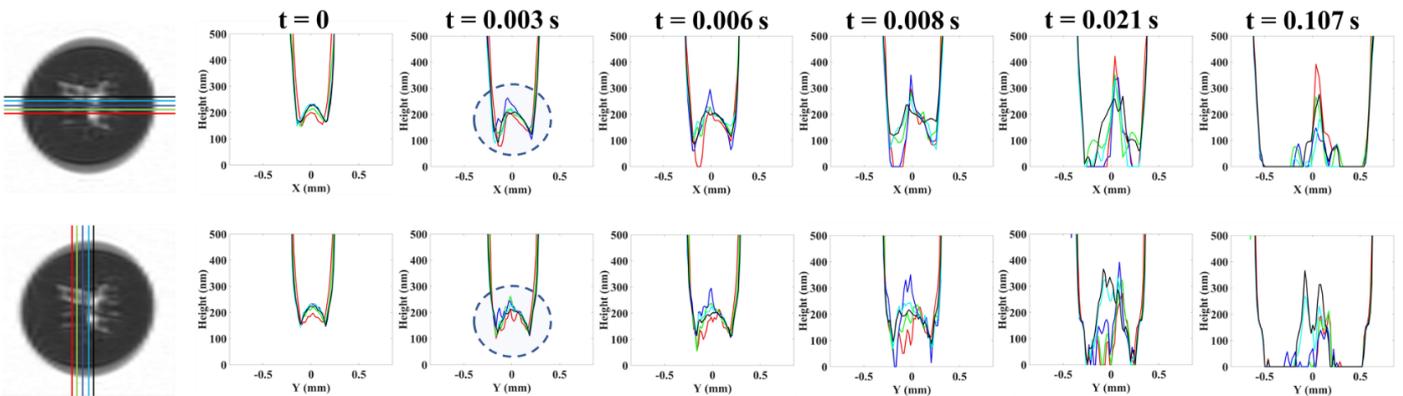

**Fig. S4.** 2D line heightmaps along x and y direction for PDMS/glass (A) and PDMS/OTS-*t*-glass (B) contact. Last frame for both the cases is at the instance where the motor stopped. Dashed blue circle in part (B) highlight the instability in elastohydrodynamic deformation. Five heightmaps along each x and y is shown for the corresponding positions marked in gray-scale images. All the cases are considered around the center as the initial projected apparent area for elastohydrodynamically deformed lens is small.

**SI, Section 5. Spatiotemporal heightmaps and corresponding raw images of different states during contact formation and water evacuation from a biologically soft human thumb/hard glass contact.**

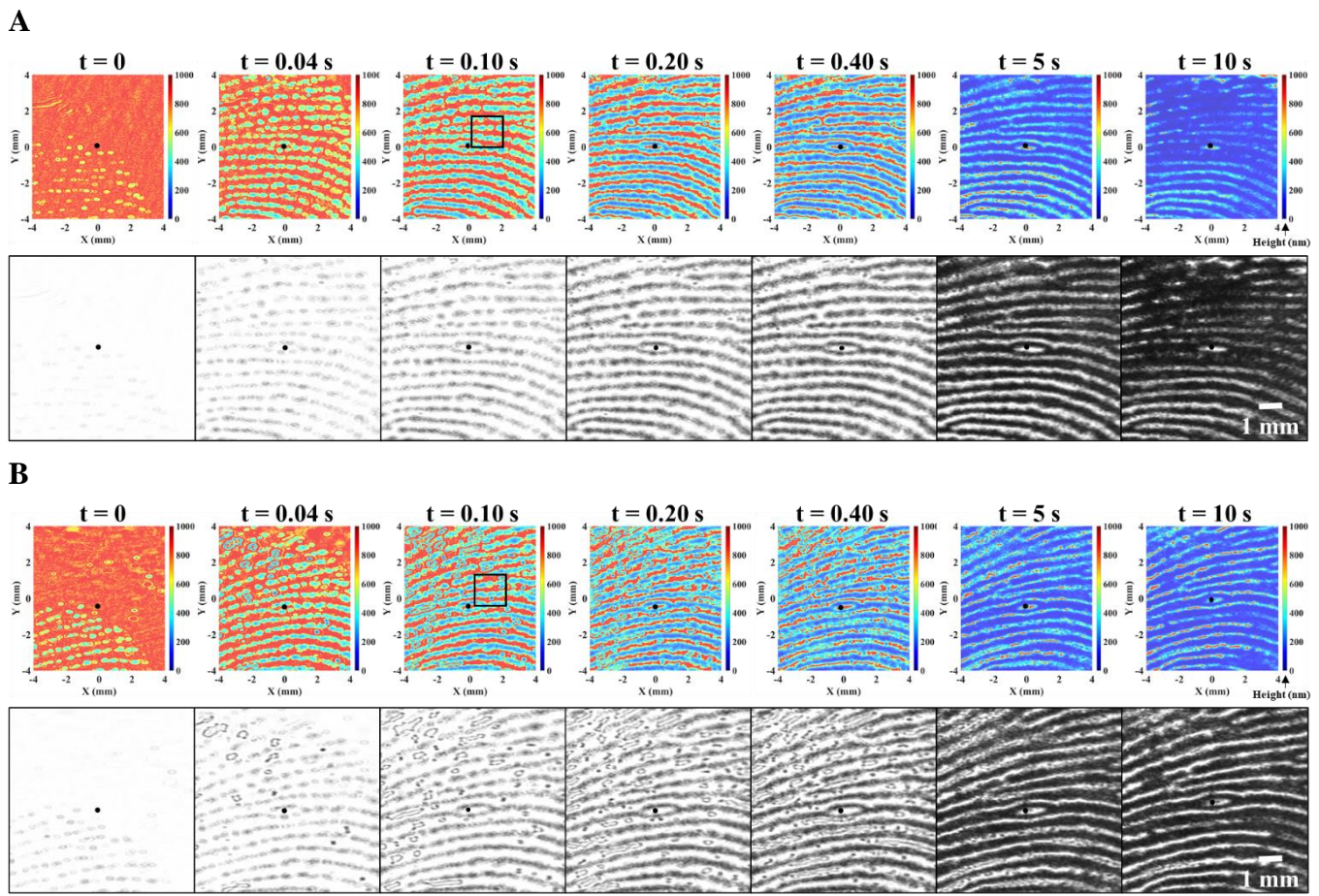

**Fig. S5.** Spatiotemporal height maps and corresponding gray-scale images for a human thumb in contact with glass (A) and OTS-*t*-glass (B) capturing various events while contact formation and water evacuation. Black solid dot in all the images act as a reference point. Both the contacts are loaded with a normal force of 10 N. Some of the pixels (white region) in these height maps are saturated due to the upper limit (Methods) of the FTIR technique. The black solid and dashed box at t = 0.10 s represent the region 1 and 2 used for evaluating volume versus time in Fig. 4 (main text).

**SI, Section 6.** Log(K) versus cosine of substrate-water contact angle plots for three moduli of PDMS lens showing similar transition behaviour as observed in, Fig. 2, main manuscript.

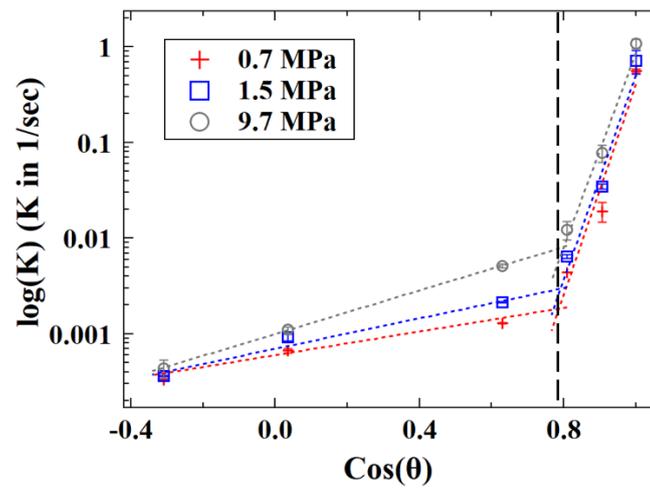

**Fig. S6.** Log(K) versus cosine of substrate water contact angle for three moduli of PDMS lens showing the transition at $\text{Cos}^{-1}(0.78)$ ($\theta \approx 38°$).

**SI, Section 7.** Raw videos for the biologically soft human thumb/hard glass contact showing the events such as contact formation and water evacuation (will be shared with the upcoming Journal version of this article).
Thumb/Glass Contact: Video S7
Thumb/OTS-*t*-Glass Contact: Video S8